\documentclass[
    reprint,
    aps,
    prapplied,
    amsmath,
    amssymb,
    superscriptaddress,
    longbibliography,
    english
]{revtex4-2}

\usepackage{graphicx}   
\usepackage{physics}
\usepackage{siunitx}
\usepackage{wasysym}
\usepackage{float}
\usepackage{color}
\usepackage{bbold}
\usepackage{amsmath}
\usepackage{amsfonts}
\usepackage{amssymb}
\usepackage[title]{appendix}

\usepackage[hidelinks,colorlinks=true,linkcolor=blue,filecolor=blue,urlcolor=blue,citecolor=blue]{hyperref}
\usepackage[all]{hypcap}
\newcommand*\diff{\mathop{}\!\mathrm{d}}
\usepackage{mathtools}

\begin{document}

    \title{Controlled-Controlled-Phase Gates for Superconducting Qubits \\ Mediated by a Shared Tunable Coupler}

    \author{Niklas J. Glaser}
    \email{niklas.glaser@wmi.badw.de}
    \affiliation{Physik-Department, Technische Universität München, 85748 Garching, Germany}
    \affiliation{Walther-Mei{\ss}ner-Institut, Bayerische Akademie der Wissenschaften, 85748 Garching, Germany}
    \author{Federico Roy}
    \affiliation{Walther-Mei{\ss}ner-Institut, Bayerische Akademie der Wissenschaften, 85748 Garching, Germany}
    \affiliation{Theoretical Physics, Saarland University, 66123 Saarbr\"ucken, Germany}

    \author{Stefan Filipp}
    \email{stefan.filipp@wmi.badw.de}
    \affiliation{Physik-Department, Technische Universität München, 85748 Garching, Germany}
    \affiliation{Walther-Mei{\ss}ner-Institut, Bayerische Akademie der Wissenschaften, 85748 Garching, Germany}
    \affiliation{Munich Center for Quantum Science and Technology (MCQST), Schellingstra\ss e 4, 80799 München, Germany}

    \begin{abstract}
        Applications for noisy intermediate-scale quantum computing devices rely on the efficient entanglement of many qubits to reach a potential quantum advantage.
        Although entanglement is typically generated using two-qubit gates, direct control of strong multi-qubit interactions can improve the efficiency of the process.
        Here, we investigate a system of three superconducting transmon-type qubits coupled via a single flux-tunable coupler.
        Tuning the frequency of the coupler by adiabatic flux pulses enables us to control the conditional energy shifts between the qubits and directly realize multi-qubit interactions.
        To accurately adjust the resulting controlled relative phases, we describe a gate protocol involving refocusing pulses and adjustable interaction times.
        This enables the implementation of the full family of pairwise controlled-phase (CPHASE) and controlled-controlled-phase (CCPHASE) gates.
        Numerical simulations result in fidelities around \SI{99}{\percent} and gate times below \SI{300}{\nano\second} using currently achievable system parameters and decoherence rates.
    \end{abstract}
    \maketitle

    \section{Introduction}\label{sec:introduction}

    Superconducting quantum circuits with fast and high-fidelity single- and two-qubit operations are considered promising candidates for quantum applications~\cite{Kjaergaard2020}.
    Recent experiments on quantum processors with dozens of superconducting qubits demonstrate the maturity of this platform~\cite{Arute2019, Jurcevic2021, Mooney2021, Gong2021} and move this technology well into the era of noisy intermediate-scale quantum (NISQ) devices~\cite{Preskill2018}.
    Promising NISQ algorithms, such as the variational quantum eigensolver (VQE)~\cite{Kandala2017, Peruzzo2014, Moll2018, Googleaiquantum2020, Ganzhorn2019} for quantum chemistry simulation and the quantum approximate optimization algorithm (QAOA)~\cite{Lacroix2020, Hill2021, Farhi2014, Harrigan2021} for complex optimization tasks, provide useful applications that do not require error correction.
    However, these rely on the efficient entanglement of many qubits to ensure that the coherence properties of the quantum system survive over the runtime of the algorithm.

    The standard approach for generating multi-qubit entanglement is the digital decomposition of multi-qubit operations into a discrete set of native single- and two-qubit gates~\cite{Barenco1995, Vartiainen2004, Shende2009, Shi2002}.
    This comes at the cost of a substantial overhead in runtime and qubit number, in particular if the connectivity between qubits is low.
    In superconducting qubit architectures, this has led to the development of problem-specific continuous two-qubit gate sets to partially reduce the overhead~\cite{Ganzhorn2019, Lacroix2020, Foxen2020, Abrams2020}.
    A complementary strategy is the direct use of multi-qubit entangling operations, which can significantly enhance the efficiency to create large-scale entanglement.
    This strategy is employed in trapped ion systems with convincing demonstrations using common vibrational modes~\cite{Molmer1999,  Kielpinski2002, Kranzl2022} and proposals using cavity-mediated interactions~\cite{Sorensen2003, Ramette2022}.
    The challenge is, however, to design multi-qubit operations that are fast and accurate when compared to an equivalent two-qubit gate decomposition, to avoid decoherence and reach high fidelities.
    For superconducting qubit architectures, several techniques for implementing multi-qubit operations have been investigated, e.g.,     utilizing non-computational qutrit states to implement more efficient digital decompositions~\cite{Mariantoni2011, Fedorov2012, Hill2021, Nikolaeva2022, Chu2021_QuAND}, applying simultaneous pairwise couplings to generate effective multi-qubit interactions~\mbox{\cite{Kim2022, Baker2021, Gu2021, Nagele2022, Zhang2022}} or introducing a shared coupling element acting as a multi-qubit coupler~\cite{Mezzacapo2014, Paik2016, Song2017, Lu2022, Song2019, Menke2022}.
    Each approach comes with its specific advantages and challenges.
    For instance, implementations based on qutrit decomposition are easy to calibrate since they use existing gate types but may suffer from decoherence due to long periods spent in higher excited states.
    Simultaneous pairwise couplings can be implemented on existing architectures but may be limited by low effective interaction strengths and spurious qubit interactions.
    And shared coupling elements provide high connectivity, but exhibit unwanted interaction terms caused by frequency crowding.

    In this work we investigate a system of three transmon-type qubits coupled via a shared transmon-type coupler, which is tunable in frequency via an external flux.
    Previous work has shown that two-qubit CPHASE gates can be realized by controlling the frequency of the tunable coupler~\cite{Yan2018, Collodo2020, Xu2020, Chu2021, Stehlik2021}.
    Here, we extend this scheme to a three-qubit system, and demonstrate that two- and three-body interaction terms can be utilized to implement the full family of controlled-controlled-phase (CCPHASE) and simultaneous pairwise controlled-phase (CPHASE) gates.
    These effective interactions are activated by adiabatic flux pulses and originate from conditional energy shifts due to hybridizations of qubit and coupler states within the same excitation manifold.
    To control the acquired two- and three-qubit phases we devise a flexible refocusing scheme.
    Therefore, strong couplings between the qubits and the tunable coupler are utilized to realize fast entangling gates with a low population in higher excited states and a large on-off coupling ratio.

    \section{Description of the system}\label{sec:the-system}

    \begin{figure}[t]
        \centering
        \includegraphics[width=0.75\columnwidth]{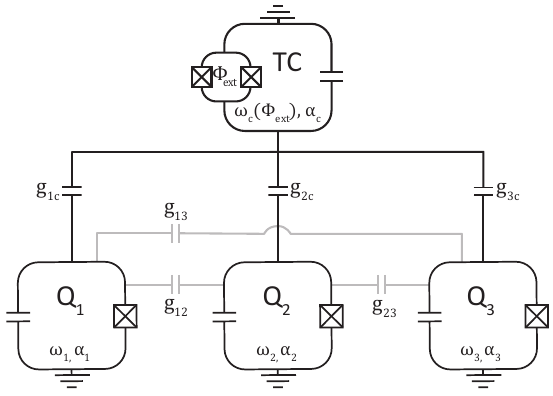}
        \caption{
            Circuit representation of the fixed-frequency transmon-type qubit setup. The qubits Q$_1$, Q$_2$ and Q$_3$ with frequencies $\omega_i$ and anharmonicity $\alpha_i$ are capacitively coupled
            to a flux-($\Phi_\text{ext}$)-tunable coupler (TC) with frequency $\omega_c$ and anharmonicity $\alpha_c$.
            The couplings $g_{ic}$ between the qubits and the coupler are assumed to be larger than the stray capacitive couplings $g_{ij}$ between qubits (gray lines).
        }
        \label{fig:3Q_architecture}
    \end{figure}

    \begin{figure}[t]
        \includegraphics[width=0.95\linewidth,height=0.5\textheight,keepaspectratio=True]{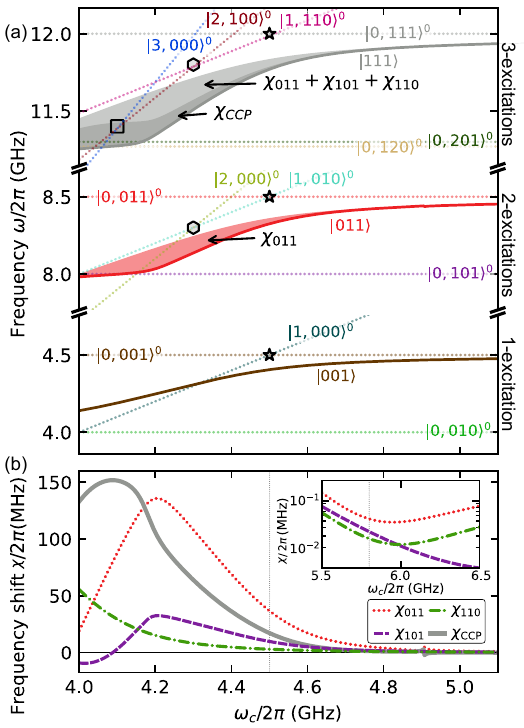}
        \caption{
            (a) Energy-level diagram for selected states of the three-qubit system. Other states are discussed in Appendix~\ref{app:full_energies}. The computational adiabatic states $\ket{n_1 n_2 n_3}$ (solid lines) and the bare states $\ket{n_c, n_1 n_2 n_3}^0$ (dotted lines) are shown in the \mbox{single-,} two- and three-excitation manifolds as a function of $\omega_c$. The frequencies $\omega$ of the states are given relative to the ground state $\ket{0,000}^0$.
            Avoided crossings relevant to the adiabatic states are highlighted:
            $\ket{1, n_1 n_2 0}^0 \leftrightarrow \ket{0, n_1 n_2 1}^0$ (stars), $\ket{2, n_1 0 0}^0 \leftrightarrow \ket{1, n_1 1 0}^0$ (hexagons) and $\ket{3, 0 0 0}^0 \leftrightarrow \ket{2, 1 0 0}^0$ (square).
            The red shaded area represents the shift $\chi_{011}$ of $\ket{011}$ from the sum of the frequencies of the respective single-excitation states.
            The grey shaded areas represent the equivalent shift $\chi_{111}$ of $\ket{111}$ with contributions from two-qubit terms (light grey) and from a three-qubit term $\chi_\text{CCP}$ (dark grey). The frequency shifts are calculated according to Eq.~\eqref{eq:frequency_shifts}, after numerical diagonalization.
            (b) Energy shifts $\chi$ as a function of $\omega_c$. Pink, red and purple solid lines show two-qubit frequency shifts and the gray line shows the three-qubit frequency shift $\chi_\text{CCP}$. Light grey vertical line shows the operating point of the coupler. The inset shows the frequency shifts near the idling point at \SI{5.8}{\giga\hertz} (vertical line) on a logarithmic scale. $\chi_\text{CCP}$ lies below the displayed range.
        }
        \label{fig:minimal_eigenergies}
    \end{figure}

    We consider three fixed-frequency transmon-type~\cite{Koch2007} qubits (Q$_1$, Q$_2$, Q$_3$), coupled via a frequency-tunable coupler (TC) \cite{Mckay2016, Yan2018, Sung2021, Chen2014} [see Fig.~\ref{fig:3Q_architecture}] described by the Hamiltonian
    \begin{align}
        \begin{aligned}
            H_\text{sys}   & = H_0 + H_{\text{int}} + H_\text{drive} \\
            H_0          & = \sum_{i=1, 2, 3} \omega_i \hat{a}_i^\dagger \hat{a}_i + \frac{\alpha_i}{2} \hat{a}_i^\dagger \hat{a}_i^\dagger \hat{a}_i \hat{a}_i
            \\&+ \omega_c(\Phi_\text{ext}) \hat{a}_c^\dagger \hat{a}_c + \frac{\alpha_c}{2} \hat{a}_c^\dagger \hat{a}_c^\dagger \hat{a}_c \hat{a}_c \\
            H_{\text{int}}   & = \frac{1}{2}\sum_{i \in \{1,2,3,c\}}\sum_{j \neq i} g_{ij}  (\hat{a}_i^\dag - \hat{a}_i)(\hat{a}_j^\dag - \hat{a}_j)                \\
            H_\text{drive} & = \sum_{i=1,2,3} \Omega_i(t) (\hat{a}_i^\dagger + \hat{a}_i),
            \label{eq:3Q_sys_Hamiltonian}
        \end{aligned}
    \end{align}
    where we set $\hbar = 1$.
    $H_0$ is the Hamiltonian of the bare uncoupled system with creation (annihilation) operators $\hat{a}_i^\dagger$ ($\hat{a}_i$), frequencies $\omega_i$ and anharmonicities $\alpha_i$ of the qubits and the coupler ($i=1,2,3,c$).
    The coupler frequency $\omega_c$ can be tuned by applying an external flux $\Phi_\text{ext}$.
    In the following we assume the experimentally realizable values $\omega_i/2\pi =\{3.5,4,4.5,4.5-5.8\}~\textrm{GHz}$ and $\alpha_i/2\pi = \{-200,-230,-200,-300\}~\text{MHz}$.
    We define the bare basis with the eigenstates $\ket{n_c, n_1 n_2 n_3}^0$ of $H_0$, where $n_i$ is the excitation number of element $i$.
    The interaction Hamiltonian $H_\text{int}$ models the capacitive couplings between the elements, for which we choose $g_{ic}/2\pi = \{150,150,120\}~\text{MHz}$ between qubits Q$_i$ and the tunable coupler TC and stray direct qubit-qubit couplings $g_{i,i+1}/2\pi = \{13,14,10\}~\text{MHz}$, which are an order of magnitude smaller.
    The term $H_\text{drive}$  models microwave drives on qubits Q$_i$ with respective Rabi rates $\Omega_i(t)$.
    At idle times and for single-qubit operations the coupler frequency $\omega_{c,\text{idle}}/ 2\pi = \SI{5.8}{\giga\hertz}$ is chosen to lie above the qubit frequencies where the effective coupling between the qubits is minimized~\cite{sete_2021a}.
    Tuning the coupler frequency close to the qubit frequencies leads to energy shifts and enhanced interaction strengths between the qubits~\cite{Yan2018}, as shown in the energy-level diagram in Fig.~\ref{fig:minimal_eigenergies}(a).
    To describe the protocol based on adiabatically modifying the frequency of the tunable coupler, we use the notation $\ket{n_1 n_2 n_3}(\omega_{c})$ for an instantaneous eigenstate of the system within the $(n_1+n_2+n_3)$-excitation manifold which is adiabatically connected to the initial state $\ket{n_1 n_2 n_3}(\omega_{c,\text{idle}}) \approx \ket{0, n_1 n_2 n_3}^0$ with zero excitations in the coupler.
    Since we are interested only in the dynamics in the qubit subspace, we diagonalize the Hamiltonian in Eq.~\eqref{eq:3Q_sys_Hamiltonian} and restrict ourselves to the computational qubit states with zero excitations in the coupler.
    This results in the Hamiltonian
    \begin{equation}
        \begin{aligned}
            \tilde{H}_\text{comp} & = \sum_{n_1, n_2, n_3\in\{0,1\}} \tilde{\omega}_{n_1 n_2 n_3} \ket{n_1 n_2 n_3}\bra{n_1 n_2 n_3},
            \label{eq:comp_Hamiltonian}
        \end{aligned}
    \end{equation}
    where $\tilde{\omega}_{n_1 n_2 n_3}$ are the instantaneous eigenfrequencies of the adiabatic states $\ket{n_1 n_2 n_3}$.

    Tuning the coupler frequency close to the qubit frequencies results in avoided crossings and energy shifts of the adiabatic states.
    The first relevant avoided crossings occur when the tunable coupler and the qubit with the highest transition frequency (Q$_3$) become resonant ($\omega_c = \omega_3$).
    These crossings, denoted by stars ($\star$) in Fig.~\ref{fig:minimal_eigenergies}, occur for all adiabatic states with one excitation in Q$_3$, i.e., $\ket{001},\ket{011},\ket{101}\text{(not shown)}$ and $\ket{111}$.
    Lowering the coupler frequency further, leads to a hybridization and an energy shift $\chi_{011}$ on $\ket{011}$ when the second excited state of the coupler $\ket{2,000}^0$ and the state $\ket{1,010}^0$ become resonant ($\omega_c =\omega_2 - \alpha_c$), as denoted by the hexagon ($\hexagon$) in Fig.~\ref{fig:minimal_eigenergies}.
    Similarly, the state $\ket{111}$ in the three-excitation manifold is shifted in energy by $\chi_{011}$ due to the $\ket{2,100}^0 \leftrightarrow \ket{1,110}^0$ interaction.
    These additional energy shifts introduce a two-body interaction term between qubits Q$_2$ and Q$_3$.
    Finally, the avoided crossing of the bare states $\ket{3,000}^0$ and $\ket{2,100}^0$, denoted by a square ($\square$) in Fig.~\ref{fig:minimal_eigenergies},
    leads to a hybridization of the adiabatic state $\ket{111}$.
    The resulting energy shift $\chi_\text{CCP}$ introduces the targeted three-body interaction for implementing controlled-controlled-phase gates.
    We note that a negative coupler anharmonicity is required for the optimal succession of the avoided crossings for the generation of the discussed energy shifts.
    The relevant two- and three-qubit energy shifts are determined from the diagonalized Hamiltonian as
    \begin{align}
        \begin{aligned}
            \chi_{011} &=  \tilde{\omega}_{011} - (\tilde{\omega}_{001} + \tilde{\omega}_{010}) \\
            \chi_{101} &=  \tilde{\omega}_{101} - (\tilde{\omega}_{001} + \tilde{\omega}_{100}) \\
            \chi_{110} &=  \tilde{\omega}_{110} - (\tilde{\omega}_{010} + \tilde{\omega}_{100}) \\
            \chi_\text{CCP} &=  \tilde{\omega}_{111} - (\tilde{\omega}_{001} + \tilde{\omega}_{010} + \tilde{\omega}_{100}) \\ &\qquad- (\chi_{011} + \chi_{101} + \chi_{110}).
        \end{aligned}
        \label{eq:frequency_shifts}
    \end{align}
    Tuning the coupler frequency close to the aforementioned avoided crossings results in large shifts of $\chi_\text{CCP}>\SI{150}{\mega\hertz}$ and $\chi_{011}>\SI{120}{\mega\hertz}$ [see Fig.~\ref{fig:minimal_eigenergies}(b)].
    We choose the coupler frequency for gate operation at $\omega_c^{\text{op}}/ 2\pi = \SI{4.5}{\giga\hertz}$ to balance the magnitude of the energy shifts and the population losses due to non-adiabatic effects of the flux pulses, however, the optimal working points for idling and gate operation depend on the specific Hamiltonian parameters.
    
    At the idling position $\omega_{c, \text{idle}}$ only minimal energy shifts, below $\SI{80}{\kilo\hertz}$, arise [see inset of Fig.~\ref{fig:minimal_eigenergies}(b)].

    To describe the entangling action of the Hamiltonian $\tilde{H}$, we switch to a frame rotating at the individual qubit frequencies
    \begin{equation}
        \begin{aligned}
            \tilde{H}= \chi_{011} H_{23} + \chi_{101} H_{13} + \chi_{110} H_{12} + \chi_\text{CCP} H_\text{CCP},
            \label{eq:cphase_Hamiltonian}
        \end{aligned}
    \end{equation}
    with the two-qubit CPHASE$_{kl}$ terms $H_{kl} = \ket{11}_{kl}\bra{11}_{kl}$,
    acting on the qubit subspace Q$_k$ and Q$_l$, and a three-qubit CCPHASE term $H_\text{CCP} = \ket{111}\bra{111}$.
    In experiments this frame rotation corresponds to applying single-qubit Z-phase gates~\cite{Mckay2017}.
    In this frame, an evolution under $\tilde{H}$ corresponds to a combination of controlled-phase (CPHASE) and controlled-controlled-phase (CCPHASE) gates
    \begin{align}
        \begin{aligned}
            U(\phi_{011},\phi_{101},\phi_{110},\phi_\text{CCP}) =
            \exp(- i \int_0^{\tau} \tilde{H}(t) \diff t) \\
            \quad\,\, = \text{CPHASE}_{12}(\phi_{110})\times \text{CPHASE}_{13}(\phi_{101})\times \\
            \quad\quad\text{CPHASE}_{23}(\phi_{011})\times \text{CCPHASE}(\phi_\text{CCP}),
        \end{aligned}
    \end{align}
    with the entangling phases
    \begin{align}
        \label{eq:cond_phases}
        \phi_j = - \int_0^{\tau} \chi_j \diff t &  & \text{for } j\in\{001,101,110,\text{\small CCP}\}.
    \end{align}

    \section{Multi-qubit gates by Hamiltonian refocusing}\label{sec:hamiltonian-refocusing}

    To implement a pure CCPHASE$(\phi_\text{CCP})$ gate without two-qubit CPHASE contributions the conditions \mbox{$\phi_{011}=k\cdot2\pi,\ \phi_{101} = l\cdot2\pi,\ \phi_{110} = m\cdot2\pi$} for Eq.~\eqref{eq:cond_phases} need to be fulfilled with integers $k,l,m$.
    However, using only the external flux applied on the coupler as a control parameter the gate duration is bounded from below by the time needed to accumulate at least a phase of $2\pi$ on all states.
    Additionally, matching all phases may require complex trajectories or a greater time overhead.
    These problems can be resolved by utilizing a refocusing scheme~\cite{Meiboom1958, Haeberlen1968, Brinkmann2016}, where the evolution under $\tilde{H}$ is interleaved with single-qubit $\pi$-pulses to permute the states accumulating the entangling phases.
    Effective two- and three-qubit gate Hamiltonians such as
    \begin{align}
        \label{eq:wanted_eff_ham}
        H_\text{eff}^{kl}=\chi_\text{eff}H_{kl} &  & \text{and} &  & H_\text{eff}^\text{CCP}=\chi_\text{eff}H_\text{CCP}
    \end{align}
    can then be realized by choosing interaction times such that unwanted phase contributions cancel out, as discussed in the following.

    \subsection{Controlled-phase gate}

    \begin{figure}[t]
        \includegraphics[width=1\linewidth,height=0.6\textheight,keepaspectratio=True]{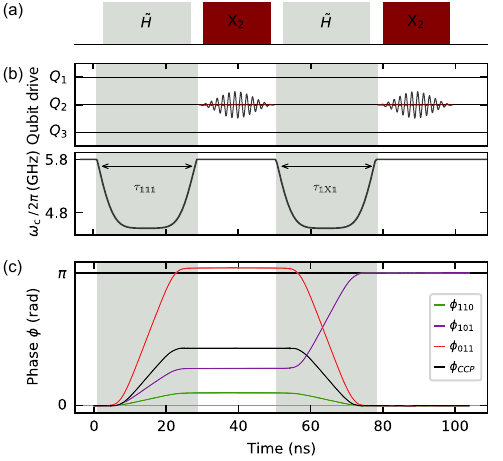}
        \caption{
            Numerical simulation of a CPHASE$_{13}(\pi)$ gate between Q$_1$ and Q$_3$.
            (a) The interaction Hamiltonian $\tilde{H}$ is applied twice, interleaved by single-qubit $\pi$-pulses on Q$_2$.
            (b) Pulse sequence on Q$_2$ and the tunable coupler. Interactions are activated by tuning the coupler frequency $\omega_c$, with \SI{27}{\nano\second} long pulses. Single-qubit $\pi$-rotations are implemented on $Q_2$ by \SI{20}{\nano\second} long microwave pulses.
            (c) Instantaneous entangling phases during the pulse sequence. The final value of the collected phase of $\phi_{110}, \phi_{011}$ and $\phi_\text{CCP}$ is zero, while the value of $\phi_{101}$ is $\pi$, which realizes a CPHASE gate between Q$_1$ and Q$_3$.
        }
        \label{fig:figure_CPHASE}
    \end{figure}
    
    \begin{table}[t]
        \centering
        \includegraphics[width=0.9\linewidth]{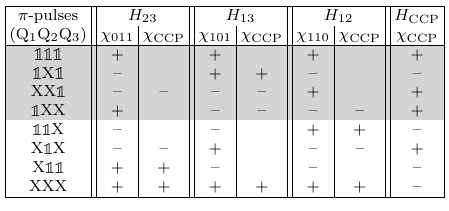}
        \caption{
            Sign of the coefficients of the effective Hamiltonians in Eq.~\eqref{eq:cphase_Hamiltonian} after applying leading and trailing $\pi$-pulses on the respective qubits denoted by an X ($\mathbb{1}$ denotes no pulse).
            Each Hamiltonian term $H_{kl}, H_\text{CCP}$ has positive or negative coefficients from any of the conditional shifts $\chi_{011},\chi_{101},\chi_{110}, \chi_{\text{CCP}}$, empty cells denote vanishing coefficients.
            The implementation of the proposed CCPHASE gate makes use of the first four permutations in the table (light grey).
        }
        \label{tab:permutated_shifted_freqs}
    \end{table}

    To generate a two-qubit CPHASE gate the non-interacting qubit needs to be dynamically decoupled~\mbox{\cite{Hahn1950,Carr1954}} from the rest of the system so that only interactions between the other two qubits remain.
    As an example, we consider a CPHASE gate between Q$_1$ and Q$_3$, for which two interaction periods of equal duration $\tau$ are interleaved with two $\pi$-pulses on $Q_2$ ($X_2$) as shown in Fig.~\ref{fig:figure_CPHASE}(a).
    The two $\pi$-pulses alter the action of the second interaction Hamiltonian as
    \begin{align}
        \label{eq:refocusing}
        \begin{aligned}
            \tilde{H}^{\mathbb{1}\mathrm{X}\mathbb{1}} = & \,\, X_2\,\tilde{H}\,X_2\\
            \cong             & - \chi_{011} H_{23} + (\chi_{101}+\chi_\text{CCP}) H_{13} \\
            & - \chi_{110} H_{12} - \chi_\text{CCP} H_\text{CCP},
        \end{aligned}
    \end{align}
    where $\cong$ defines an equality that neglects global and single-qubit terms (see Appendix~\ref{app:refocusing-derivation} for further details).
    In $\tilde{H}^{\mathbb{1}\mathrm{X}\mathbb{1}}$ the sign of all components is inverted except for $H_{13}$, which increases by $\chi_\text{CCP}$.
    The action of the full CPHASE sequence is then given by
    \begin{align}
        \begin{aligned}
            &X_2 e^{-i\tau\tilde{H}} X_2 e^{-i\tau\tilde{H}}
            =e^{-i\tau[\tilde{H}^{\mathbb{1}\mathrm{X}\mathbb{1}} + \tilde{H}]}
            \\&=e^{-i\tau[2\chi_{101}+\chi_\text{CCP}]H_{13}}
            =\text{CPHASE}_{13}(\phi_{101}),
        \end{aligned}
    \end{align}
    where we have used the equality $U e^{- i \tilde{H}} U^{-1} = e^{-i U\tilde{H}U^{-1}}$ and that the interaction Hamiltonians are diagonal and, hence, commute with each other.
    This is equivalent to evolving for $2\tau$ under the effective Hamiltonian of Eq.~\eqref{eq:wanted_eff_ham} with $\chi_\text{eff}=\chi_{101}+\chi_\text{CCP}/2$ and $\phi_{101}=\tau[2\chi_{101}+\chi_\text{CCP}]$.

    To verify the performance of the CPHASE gate we numerically simulate the CPHASE$_{13}(\pi)$ gate resulting in a fidelity of $99.77\%$ at a gate time of approximately \SI{100}{\nano\second} (see Appendix~\ref{app:simulation} for simulation details), assuming that decoherence times are much longer than the gate times.
    To tune the coupler from its idle position~$\omega_c^\text{id}=\SI{5.8}{\giga\hertz}$ to the operation point $\omega_c^\text{op}=\SI{4.5}{\giga\hertz}$ we use rectangular pulses with Gaussian edges:
    \begin{equation}
        \label{eq:half_Gaussian_square_pulse}
        \omega_c(t) = \omega_c^\text{id} + (\omega_c^\text{op}-\omega_c^\text{id}) \erf\left(\frac{t}{\tau_R}\right) \erf\left(\frac{\tau-t}{\tau_R}\right),
    \end{equation}
    where $\tau=\SI{27}{\nano\second}$ is the width of the pulse and $\erf$ is the Gaussian error function with a rise time characterized by $\tau_R=\SI{5}{\nano\second}$ [Fig.~\ref{fig:figure_CPHASE}(b)].
    The $\pi$-pulses are driven by a microwave pulse, with a \SI{20}{\nano\second} DRAG-corrected Gaussian envelope~\cite{Motzoi2009}, with fidelities $> \SI{99.96}{\percent}$.
    The numerically simulated evolution of the entangling phases is shown in Fig.~\ref{fig:figure_CPHASE}(c).
    Initially all entangling phases evolve at a positive rate.
    However, after the $\pi$-pulse on Q$_2$ all accumulation rates are inverted except for $\phi_{110}$, resulting in the desired entangling phases of $\phi_{110}=\phi_{011}=\phi_\text{CCP} = 0$ and $\phi_{101} = \pi$.
    With this approach we can realize all pairwise two-qubit CPHASE$_{kl}$($\phi_{kl}$) gates by performing interleaved $\pi$-pulses on the qubit not participating in the gate.
    The effective Hamiltonians for the different combinations of enclosing $\pi$-pulses are listed in Tab.~\ref{tab:permutated_shifted_freqs}.

    \subsection{Controlled-controlled-phase gate}
\begin{figure}[b]
        \centering
        \includegraphics[width=1\linewidth,height=0.6\textheight,keepaspectratio=True]{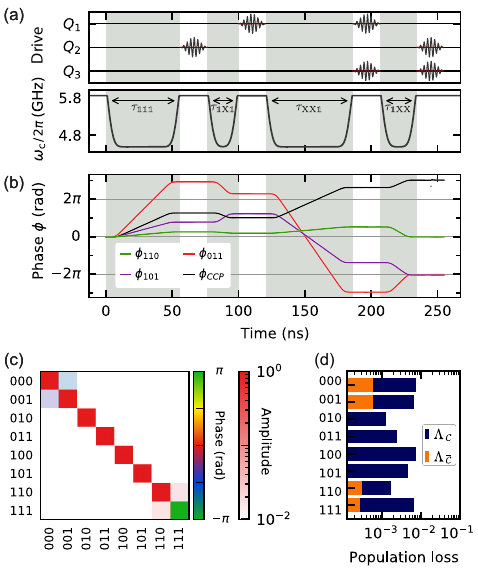}
        \caption{
            Numerical simulation of a CCPHASE$(\pi)$ gate. (a)
            Four independent adiabatic pulses modulate the frequency $\omega_c$ of the coupler, with durations  $\tau_{\mathbb{1}\mathbb{1}\mathbb{1}},\tau_{\mathbb{1}\mathrm{X}\mathbb{1}},\tau_{\mathrm{X}\mathrm{X}\mathbb{1}},\tau_{\mathbb{1}\mathrm{X}\mathrm{X}}$. The coupler pulses are interleaved by single-qubit refocusing pulses, on the respective qubits Q$_i$.
            (b) Instantaneous entangling phases $\phi_i$ collected during the gate execution.
            (c) Propagator of the simulated gate. Phase and amplitude of matrix elements are represented in color and opacity, respectively.
            (d) Population loss from the initialized computational states after the gate execution to other computational states $\Lambda_\mathcal{C}$ (orange), equivalent to the sum of the squared off-diagonal elements in (c), or to states outside the computational basis $\Lambda_{\bar{\mathcal{C}}}$ (blue).
        }
        \label{fig:ccphase_gate}
    \end{figure}

    To implement a CCPHASE gate the refocusing technique can be extended to four interaction periods interleaved with single-qubit $\pi$-pulses on different qubits, as shown in Fig.~\ref{fig:ccphase_gate}(a).
    For the observed relation, 

    \onecolumngrid
    $\chi_{101} > \chi_\text{CCP} > \chi_{110}$, we choose the permutations $\mathbb{1}\mathbb{1}\mathbb{1}$, $\mathbb{1}\mathrm{X}\mathbb{1}$, $\mathrm{X}\mathrm{X}\mathbb{1}$ and $\mathbb{1}\mathrm{X}\mathrm{X}$.
    We skip repeated single-qubit $\pi$-rotations which appear between permutations and thus apply six single qubit gates in total.
    To find the correct duration of each flux pulse we solve the system of equations
    
    \begin{equation}
        \label{eq:full_ccphase_lin_equation}
        \begin{pmatrix}
            \phi_{23} + k_{23} 2\pi \\
            \phi_{13} + k_{13} 2\pi \\
            \phi_{12} + k_{12} 2\pi \\
            \phi_\text{CCP} + k 2\pi
        \end{pmatrix}
        =
        \begin{pmatrix}
            \chi_{011} & - \chi_{011} & - \chi_{011} - \chi_\text{CCP} & \chi_{011}
            \\
            \chi_{101} & \chi_{101} + \chi_\text{CCP} & - \chi_{101} - \chi_\text{CCP} & - \chi_{101} - \chi_\text{CCP}
            \\
            \chi_{110} & - \chi_{110} & \chi_{110} & - \chi_{110} - \chi_\text{CCP}
            \\
            \chi_\text{CCP} & - \chi_\text{CCP} & \chi_\text{CCP} & \chi_\text{CCP}
        \end{pmatrix}
        \begin{pmatrix}
            \tau_{\mathbb{1}\mathbb{1}\mathbb{1}} \\ \tau_{\mathbb{1}\mathrm{X}\mathbb{1}} \\ \tau_{\mathrm{X}\mathrm{X}\mathbb{1}} \\ \tau_{\mathbb{1}\mathrm{X}\mathrm{X}}
        \end{pmatrix}
    \end{equation}
    \twocolumngrid
    
    \noindent derived from the Hamiltonian coefficients in Table~\ref{tab:permutated_shifted_freqs},
    assuming a linear increase of entangling phases with the duration of the flux pulses (see Appendix~\ref{app:pulse_par_opt}).
    As for the two-qubit CPHASE gate above, we simulate the coherent dynamics for a $\text{CCPHASE}(\pi)$ gate, with pulse durations (55,25,65,20)~\si{\nano\second} and plot the evolution of the entangling phases in Fig.~\ref{fig:ccphase_gate}(b).
    With these timings, we obtain final entangling phases of $\phi_{110} = 0, \phi_{011} = \phi_{101} = -2\pi$ and $\phi_\text{CCP} = 3\pi$ and a simulated fidelity of \SI{99.58}{\percent} in \SI{245}{\nano\second} including the single qubit gates [see Fig.~\ref{fig:ccphase_gate}(c)]. Leakage caused by imperfectly adiabatic pulses as the main coherent error contribution as discussed in Section~\ref{sec:error_contribution}.

    \begin{figure}[b]
        \vspace{-5mm}
        \centering
        \includegraphics[width=1\linewidth, height=0.4\textheight, keepaspectratio=True]{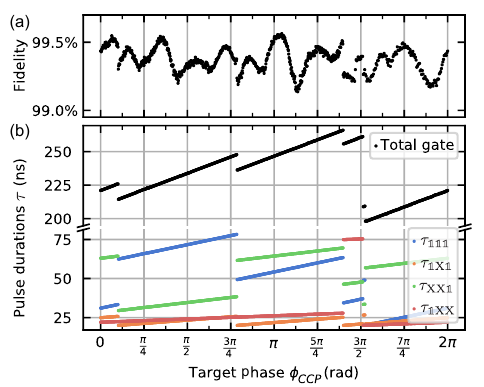}
        \caption{
            Continuous CCPHASE($\phi_\text{CCP}$) gate with varying target phase $\phi_\text{CCP}$. (a) Fidelity and (b) gate duration including single-qubit $\pi$-pulses. Individual adiabatic pulse durations $\tau_i$ for realizing the respective target phases are shown with coloured dots. The discontinuities are caused by the $2\pi$-phase degree of freedom when solving Eq.~\eqref{eq:full_ccphase_lin_equation}.
            \label{fig:gate_times_param_gate}
        }
    \end{figure}

    \subsection{Generalized controlled-controlled-phase gate}
    Applying the same refocusing scheme, Eq.~\eqref{eq:full_ccphase_lin_equation} can be used to determine pulse durations that result in arbitrary two- and three-qubit entangling phases.
    This method therefore provides full control over all entangling phases and directly allows for the implementation of the generalized three-qubit controlled-phase gate
    \begin{align}
        \begin{aligned}
            U_{3\text{Q}\phi} & (\phi_{12},\phi_{13},\phi_{23},\phi_\text{CCP}) =                            \\
            =                 & \text{CPHASE}_{12}(\phi_{12})\times \text{CPHASE}_{13}(\phi_{13})\times \\
            & \text{CPHASE}_{23}(\phi_{23})\times \text{CCPHASE}(\phi_\text{CCP}),
        \end{aligned}
    \end{align}
    which corresponds to the simultaneous application of pairwise two- and three-qubit controlled-phase gates.
    In particular, we can implement a CCPHASE gate with an arbitrary angle $\phi_\text{CCP}$.
    We numerically evaluate the gate fidelities and pulse durations for $\phi_\text{CCP}$ continuously varying between $0$ and $2\pi$, as shown in Fig.~\ref{fig:gate_times_param_gate}.
    We find that all phase combinations can be realized with total gate lengths between \SI{195}{\nano\second} and \SI{270}{\nano\second}.
    Without including decoherence, the fidelity for all implementation lies between $99.1\%-99.6\%$, with oscillations due to periodic leakage effects (see Appendix~\ref{app:pulse_par_opt} for more details).
    Similarly, $U_{3\text{Q}\phi}$ gates with arbitrary settings of both the two- and three-qubit phases $\phi_{ij}$ and $\phi_\text{CCP}$ result in gate times below \SI{300}{\nano\second} and gate fidelities above $99\%$ (not shown).

    \section{Error Contributions}\label{sec:error_contribution}

    To assess the expected performance of the gate operations, we evaluate the error contributions from coherent errors and from decoherence on the CCPHASE($\pi$) gate.
    The coherent errors of \SI{0.42}{\percent} are dominated by leakage, i.e., all population losses $\Lambda_s = \sum_{f \neq i} {\bra{f} U \ket{i}}^2$ from computational states $\ket{i}\in\mathcal{C}$ to states outside the computational subspace $\ket{f}\in s=\bar{\mathcal{C}}$ [blue bars in Fig.~\ref{fig:ccphase_gate}(d)], which are caused by imperfect adiabatic pulses.
    Transitions to other computational states ${\ket{f}\in s=\mathcal{C}\setminus \{\ket{i}\}}$ [orange bars in Fig.~\ref{fig:ccphase_gate}(d)] caused by imperfections in single-qubit gates are another coherent error source leading to off-diagonal elements in the propagator [Fig.~\ref{fig:ccphase_gate}(c)].
    For a single flux pulse most of the losses occur from the states
    $\ket{111}$ and  $\ket{011}$ (see Appendix~\ref{app:pulse_par_opt}).
    However, permuting the states with the interleaved single-qubit pulses distributes the leakage over all computational states.
    Note that the coherent errors scale with the number of flux pulses.
    For the CPHASE gate, which uses two flux pulses, coherent errors amount to \SI{0.23}{\percent} roughly a factor two smaller than CCPHASE gate, which uses four.

    In addition to leakage, decoherence of both qubits and the coupler will limit the achievable gate fidelities.
    In particular, because of the transmon-type tunable coupler charge noise may induce errors due to the hybridization in higher-excitation manifolds.
    We therefore simulate the open-system dynamics of the system by solving the time-evolution under the Lindblad master equation
    \begin{align}
        \begin{aligned}
            \dot{\rho}&=-i[H, \hat{\rho}(t)]+\sum_{k} \left(\hat{L}_{k} \hat{\rho} \hat{L}_{k}^{\dagger}-\frac{1}{2}\left\{\hat{L}_{k}^{\dagger} \hat{L}_{k}, \hat{\rho(t)}\right\}\right) \label{eq:Lindblad_Master_equation}
        \end{aligned}
    \end{align}
    with the Hamiltonian $H$, the density matrix $\rho$ and the collapse operators $\hat{L}_k$ accounting for relaxation and dephasing.
    We assume relaxation rates that are linearly increasing with excitation number
    $\Gamma_1^{(m+1,m)} = \frac{m}{T_1}$
    where $\Gamma_1^{(j,m)}$ is the decay rate from state $\ket{j}$ to $\ket{m}$, and $T_1$ is the relaxation time of the first excited state~\cite{Koch2007, Peterer2015}.
    The pure dephasing rate $\Gamma_{\phi}^{(m)} =  \Gamma_{\phi,c}^{(m)}  + \Gamma_{\phi,e}^{(m)}$ between states $\ket{m}$ and $\ket{m+1}$ is modelled as the sum of a constant term $\Gamma_{\phi,c}^{(m)} = \frac{1}{T_{\phi}}$
    and an energy-level-dependent term $\Gamma_{\phi,e}^{(m)}$ incorporating charge noise~\cite{Koch2007, Burnett2019, Rol2019}.
    We account for a charge noise dephasing rate $\Gamma_{\phi, n}^{(m)} = \pi A_n |\epsilon_m|$ given the charge noise strength $A_n$ and the approximated charge dispersion amplitude~\cite{Koch2007}
    \begin{align}
        \epsilon_{m} \simeq(-1)^{m} E_{C} \frac{2^{4 m+5}}{m !} \sqrt{\frac{2}{\pi}}\left(\frac{E_{J}}{2 E_{C}}\right)^{\frac{m}{2}+\frac{3}{4}} e^{-\sqrt{8 E_{j} / E_{C}}}. \label{eq:charge_dispersion}
    \end{align}
    We assume equal decoherence rates on all qubits and couplers. Although, flux-tunable transmons tend to have higher dephasing rates due to their sensitivity to flux noise, this can be mitigated by introducing a coupler with an asymmetric SQUID \cite{Koch2007} and choosing idling and operating points to be at the flux sweetspots.

\begin{figure}[t]
        \centering
        \includegraphics[width=\columnwidth]{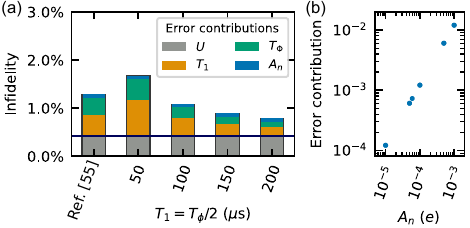}
        \caption{Effect of coherent errors and decoherence on CCPHASE gate fidelities.
            (a) The infidelity of the numerically simulated CCPHASE gate for different relaxation times $T_1$ and pure dephasing times $T_\phi$.
            The first bar shows the results of a model according to~\cite{Peterer2015}.
            Different colors denote the contributions of the respective error channels: coherent errors $U$ (grey), $T_1$ relaxation (orange), pure $T_\phi$ dephasing (green) and charge noise $A_n$ errors (blue). (b) Error contribution from charge noise for various noise strengths $A_n$.
            \label{fig:CCPHASE_decoherences}
        }
    \end{figure}

    We determine the individual contributions of each error channel by comparing the simulations with the respective error channel turned on and off.
    We find that the errors are independent of each other and can be linearly added.
    Using typical values of $T_1 = \SI{84}{\micro\second}$, $T_\phi=\SI{124}{\micro\second}$ and $A_n=\SI{6e-5}{}\,e$ (see e.g.\ Peterer et al.~\cite{Peterer2015}), we obtain a total CCPHASE$(\pi)$ gate infidelity of \SI{1.30}{\percent}, as compared to only coherent errors of \SI{0.42}{\percent}, as shown in Fig~\ref{fig:CCPHASE_decoherences}(a).
    Keeping the noise amplitude fixed we evaluate infidelities for varying values of $T_1$ and $T_\phi$, assuming that as the coherence times improve the ratio between them will remain approximately $T_1=T_\phi/2$.
    For $T_1=\SI{50}{\micro\second}$ we find that decoherence is the dominant contribution, however, for state-of-the-art coherence times $T_1 > \SI{100}{\micro\second}$~\cite{Place2021,Wang2022} decoherence and coherent errors become comparable, with total errors of approximately $\SI{1}{\percent}$ and below.
    The error caused by charge noise with a typical strength $A_n=\SI{6e-5}{}\,e$~\cite{Christensen2019, Peterer2015} is expected to be below $\SI{0.1}{\percent}$, with the error scaling roughly linearly with the charge noise amplitude, as shown in Fig.~\ref{fig:CCPHASE_decoherences}(b).

    \section{Discussion and Outlook}\label{sec:discussion-and-outlook}

    In summary, we propose a system of three qubits coupled via a shared tunable coupler and develop a pulse scheme that implements generalized three-qubit controlled-phase gates.
    Adiabatic flux pulses allow us to tune the two- and three-body interaction strengths.
    The resulting entangling phases are controlled by interleaving interaction periods with single-qubit refocusing pulses.
    With this method, we show in numerical simulations that three-qubit controlled-controlled-phase gates can be realized in less than \SI{300}{ns} with fidelities above \SI{99}{\percent} for all desired entangling phases when the effects of decoherence can be ignored.
    Taking realistic values for qubit and coupler coherence times into account we expect an added gate error below \SI{1}{\percent}.
    Recent implementations of three-qubit gates include cross-resonance-type iToffoli gates~\cite{Kim2022} with fidelities of \SI{98.3}{\percent} in 353ns, simultaneous parametric drive gates~\cite{Warren2022} with fidelities of \SI{97.9}{\percent} in 250ns, M{\o}lmer-S{\o}rensen-type gates~\cite{Lu2022} with fidelities of \SI{90.5}{\percent} in 217ns, and decomposed CCPHASE gates~\cite{Hill2021} with fidelities of \SI{87.1}{\percent} in 402ns.
    By directly utilizing the strong qubit-coupler interactions and a flexible pulse scheme, the proposed CCPHASE gate has the potential to improve both speed and fidelity as compared to these recent realizations of three-qubit gates on superconducting qubits.
    Moreover, the studied three-qubit coupler refocusing scheme allows for the implementation of pairwise controlled-phase gates with adjustable phases, thus providing greater connectivity and flexibility in comparison to two-qubit couplers.

    The proposed protocol can be modified in a number of ways to account for different experimental conditions.
    In the presence of strong charge noise the operating point of the coupler $\omega_c^\text{op}$ could be lowered, thus reducing charge noise sensitivity at the cost of longer gate times.
    Furthermore, the refocusing scheme can be adapted to result in a net-zero total applied flux, a technique known to reduce sensitivity to long-term correlated flux noise if present~\cite{Rol2019}.
    To reduce leakage errors, flux pulses could be individually optimized to harness destructive interference between multiple transitions~\cite{Shevchenko2010} and ensure local adiabaticity~\cite{Roland2002}.
    In general, a refocusing scheme similar to that presented can be applied in other superconducting qubit architectures.
    For example, in systems with two-qubit couplers and simultaneous interactions~\cite{Baker2021} it could provide greater control of all interaction terms and shorter gate durations.
    Moreover, the number of qubits connected to the coupler could be increased further, allowing for strong and controllable many-body Hamiltonians with application in variational algorithms and Hamiltonian simulations.
    Finally, in the context of quantum applications, the investigated architecture is a promising candidate for the implementation of variational algorithms designed to solve optimization problems.
    In particular, the problem Hamiltonian for MAX-3-SAT problems can be directly implemented by the generalized controlled-phase gate, providing an improvement in speed and accuracy over a gate decomposition into single- and two-qubit gates.

    \section{Acknowledgments}\label{sec:acknowledgments}

    We thank Ivan Tsitsilin, Gerhard Huber and Franz Haslbeck for insightful discussions.
    We acknowledge funding from the European Commission Marie Curie
ETN project QuSCo (Grant No. 765267), from the German Federal Ministry of Education and Research via the funding program “Quantum Technologies-From Basic Research to the Market” (project GeQCoS) under Contract No. 13N15680, and from the European FET OPEN project Quromorphic (Grant No. 828826).
    We also acknowledge funding by the Deutsche Forschungsgemeinschaft (DFG, German Research Foundation) under Project No. \mbox{FI2549/1-1}.
    We further acknowledge support by the Leibniz Supercomputing Centre, providing computing time on its \mbox{Linux-Cluster}.

    \begin{appendices}

        \section{\texorpdfstring{\\* \vspace{2mm}}~Simulations}\label{app:simulation}

        The density matrix dynamics are simulated with the q-optimize~\cite{Wittler2021} package, using time-ordered piecewise exponentiation of Hamiltonian $H_\text{sys}$ in Eq.~\eqref{eq:3Q_sys_Hamiltonian}.
        The dynamics are sampled at a rate of \SI{30}{GS\per\second}, while the control signals are sampled at \SI{2.4}{GS\per\second} with additional Gaussian filtering according to the specifications of a typically used arbitrary waveform generator, such as the HDAWG from Zurich Instruments~\cite{Zurichinstrumentsag2022}.
        We restrict the energy of the bare states to $E/(2\pi)$= $\omega_{th} / 2\pi=\SI{16}{\giga\hertz}$ and therefore take only the lowest-lying energy levels of the qubits (4,4,3 for Q$_1$,Q$_2$,Q$_3$) and the tunable coupler TC (5) into account.
        We further truncate the Hilbert space to allow only for up to four excitations in the system.

        \section{\texorpdfstring{\\* \vspace{2mm}}~Interacting energy levels}\label{app:full_energies}

        \begin{figure*}[htb]
            \centering
            \includegraphics[scale=0.9]{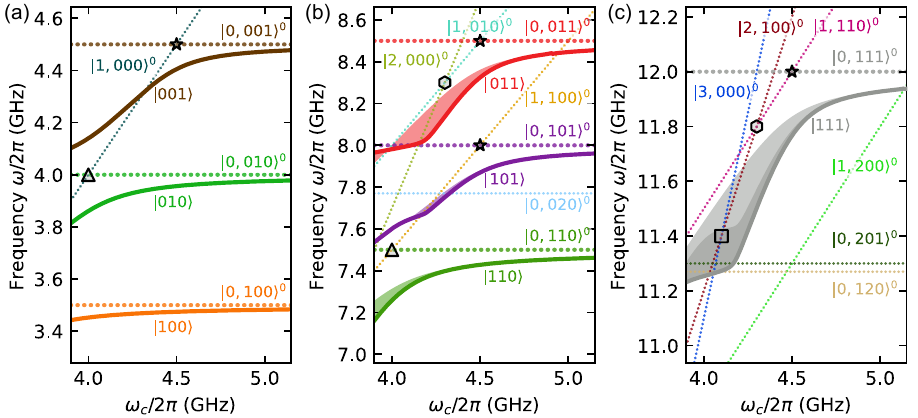}
            \caption{Energy level diagram of the three-qubit system.
            Frequencies with respect to the ground state $\ket{0,000}$ of bare states (dotted lines) and adiabatic states (solid lines) are shown as function of the coupler frequency, in the single- (a), double- (b) and triple-excitation (c) manifolds. The adiabatic states follow the eigenstates, except for avoided crossings with a gap spacing less than \SI{10}{\mega\hertz}, assuming that those avoided crossings are passed diabatically. The coupler is idling at \SI{5.8}{\giga\hertz}, defining the labelling of the adiabatic states. The relevant avoided crossings are marked by $\star$: $\ket{1, n_1 n_2 0}^0 \leftrightarrow \ket{0, n_1 n_2 1}^0$, $\hexagon$:  $\ket{2, n_1 0 0}^0 \leftrightarrow \ket{1, n_1 1 0}^0$, $\square$: $\ket{3, 0 0 0}^0 \leftrightarrow \ket{2, 1 0 0}^0$ and $\triangle$: $\ket{1, n_1 0 n_3}^0 \leftrightarrow \ket{0, n_1 1 n_3}^0$.
            \label{fig:full_energies}}
        \end{figure*}

        While in Section~\ref{sec:the-system} we discussed only the avoided crossings dominantly contributing to the energy shifts of states $\ket{011}$ and $\ket{111}$, here we extend the discussion to all crossings affecting the adiabatic computational states $\ket{n_1 n_2 n_3}, n_k \in \{0,1\}$, as shown in Fig.~\ref{fig:full_energies}.

        The energy shift for the adiabatic state $\ket{111}$ occurs due to hybridization first with the $\ket{1,110}^0$ state, then with $\ket{2,100}^0$ and finally with $\ket{3,000}^0$ [star, circle, and square, respectively, in Fig.~\ref{fig:full_energies}(c)].
        Likewise, the energy shift for the adiabatic state $\ket{011}$ occurs due to hybridization first with the $\ket{1,010}^0$ state and then with $\ket{2,000}^0$ [top star and circle, respectively, in Fig.~\ref{fig:full_energies}(b)].
        In both situations, lowering the coupler frequency $\omega_c$ past the last transition should further increase the energy shifts.
        However, this behaviour is inhibited by hybridization with other bare states, such as $\ket{0,101}^0$ for $\chi_{110}$ and $\ket{0,201}^0, \ket{0,120}^0$ for $\chi_\text{CCP}$.
        At a higher coupler frequency, instead, two further avoided crossings affect the $\ket{111}$ and $\ket{011}$ states when tuning the coupler into the interaction area: $\ket{1,200}^0\leftrightarrow\ket{0,111}^0$ and $\ket{1,100}^0\leftrightarrow\ket{0,011}^0$, respectively.
        These avoided crossings have a gap of only a few megahertz, as both are caused by four-photon transitions, and therefore need to be passed diabatically, i.e.\ with a fast passage.
        For the choice of circuit parameters it is thus essential to ensure that these four-photon crossings are located at a tunable coupler frequency $\omega_c$ above the idling point and separated from the chosen interaction area, such that the requirements on adiabaticity as well as diabaticity can both be fulfilled.

        Similarly to $\ket{011}$, the adiabatic state $\ket{101}$ would experience a conditional energy shift $\chi_{101}$ due to hybridization first with the $\ket{1,100}^0$ state [lower star in Fig.~\ref{fig:full_energies}(b)] and then with $\ket{2,000}^0$ (not shown), and the adiabatic state $\ket{110}$ would experience a conditional energy shift $\chi_{110}$ due to hybridization first with the $\ket{1,100}^0$ state [triangle in Fig.~\ref{fig:full_energies}(b)] and then with $\ket{2,000}^0$ (not shown).
        However, the interaction with $\ket{2,000}^0$ occurs at a coupler frequency $\omega_c=\omega_1 - \alpha_c$ below all qubit frequencies, making it difficult to reach areas with large energy shifts without causing leakage in at least some of the computational states.
        Even then, for the adiabatic state $\ket{101}$ a previous hybridization with the $\ket{0,020}^0$ state suppresses the energy shift $\chi_{101}$ and introduces the risk of leakage.
        Therefore, large energy shifts are only achievable on states $\ket{011}$ and $\ket{111}$.

        \section{\texorpdfstring{\\* \vspace{2mm}}~Method of refocusing}\label{app:refocusing-derivation}
        By interleaving $\pi$-pulses in the conditional phase accumulation, unwanted phase terms can be cancelled to realize CPHASE and CCPHASE gates.
        Here we consider the interaction Hamiltonian (see also Eq.~\eqref{eq:cphase_Hamiltonian} in the main text)
        \begin{equation}
            \tilde{H} = \chi_{011} H_{23} + \chi_{101} H_{13} + \chi_{110} H_{12} + \chi_\text{CCP} H_\text{CCP}.
        \end{equation}
        Applying a leading and a trailing $\pi$-X-pulse on Q$_2$ ($X_2$) results in $\tilde{H}^{\mathbb{1}\mathrm{X}\mathbb{1}}$, which can be used in conjunction with $\tilde{H}$ to realize a controlled-phase gate between Q$_1$ and Q$_3$.
        A straightforward extension of this scheme to other qubit combinations results in the effective Hamiltonians listed in Tab.~\ref{tab:permutated_shifted_freqs}.

        The effective Hamiltonian $\tilde{H}^{\mathbb{1}\mathrm{X}\mathbb{1}}$ is given by
        \begin{align}
            X_2\, e^{- i \tilde{H} \tau}\, X_2 = e^{-i X_2\,\tilde{H}\,X_2 \tau} = e^{-i \tilde{H}^{\mathbb{1}\mathrm{X}\mathbb{1}} \tau},
        \end{align}
        where the first equality is given by $X_2 X_2 = \mathbb{1}$.
        The transformed Hamiltonian terms are then given by
        \begin{align}
            \begin{aligned}
            {H}
                ^{\mathbb{1}\mathrm{X}\mathbb{1}}_{13} = X_2 H_{13} X_2 &\cong H_{13}, \\
                {H}^{\mathbb{1}\mathrm{X}\mathbb{1}}_{12} = X_2 H_{12} X_2 &\cong -H_{12}, \\
                {H}^{\mathbb{1}\mathrm{X}\mathbb{1}}_{23} = X_2 H_{23} X_2 &\cong -H_{23}, \\
                {H}^{\mathbb{1}\mathrm{X}\mathbb{1}}_\text{CCP} = X_2 H_\text{CCP} X_2 &\cong -H_\text{CCP}+H_{13}
            \end{aligned}
        \end{align}
        using the relation $X_{j}Z_{j}X_{j} = -Z_j$ and the Hamiltonian terms
        \begin{align}
            \label{eq:ham_terms}
            \begin{split}
                H_{kl} &= \ket{11}_{kl}\bra{11}_{kl} \\
                &= \left(\mathbb{1}_{k}\mathbb{1}_{l}-\mathbb{1}_{k}Z_{l}+Z_{k}\mathbb{1}_{l}+Z_{k}Z_{l}\right)/4 \\
                &\cong Z_{k}Z_{l}/4 \\
                H_\text{CCP}&=\ket{111}\bra{111}\\ &=\big(\mathbb{1}\mathbb{1}\mathbb{1}-\mathbb{1}\mathbb{1}Z-\mathbb{1}Z\mathbb{1}-Z\mathbb{1}\mathbb{1}\\&\hspace{0.8cm}+ZZ\mathbb{1}+Z\mathbb{1}Z+\mathbb{1}ZZ-ZZZ\big)/8 \\
                &\cong \left(ZZ\mathbb{1}+Z\mathbb{1}Z+\mathbb{1}ZZ-ZZZ\right)/8,
            \end{split}
        \end{align}
        written in terms of the Pauli operators $I$ and $Z$.
        Here we neglect global and single-qubit terms, denoted by $\cong$ in Eqs.~\eqref{eq:ham_terms}.
        The resulting effective Hamiltonian is then given by
        \begin{align}
            \begin{aligned}
                \tilde{H}^{\mathbb{1}\mathrm{X}\mathbb{1}} \cong
                &- \chi_{011} H_{23} + (\chi_{101}+\chi_\text{CCP}) H_{13} \\
                &- \chi_{110} H_{12} - \chi_\text{CCP} H_\text{CCP}.
            \end{aligned}
        \end{align}

        \begin{figure}[t]
            \centering
            \includegraphics[scale=1]{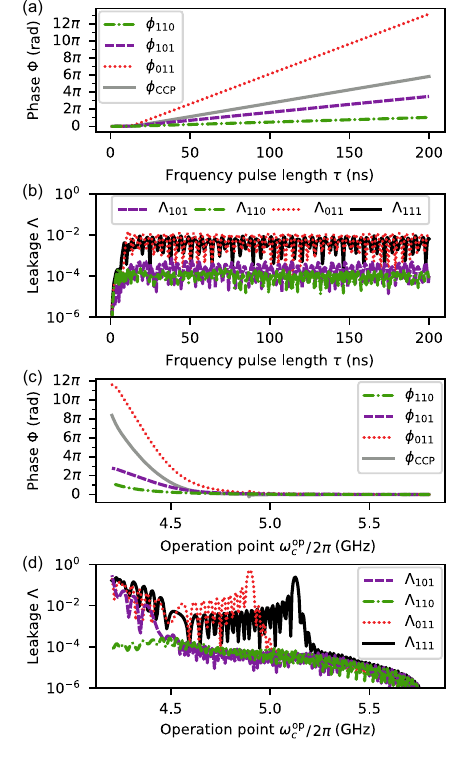}
            \caption{
                \label{fig:phase_leak_amp_t_plot}
                Simulated entangling phases $\phi_{j}$ and individual state population loss $\Lambda_{n_{1}n_{2}n_{3}}$ as a function of gate duration $\tau$ (a,b) and of coupler operating point frequency $\omega_c^{op}$ (c,d).
                For all simulations a half-Gaussian-square pulse is used, with total duration $\tau = \SI{60}{\nano\second}$, rise time $\tau_R = \SI{5}{\nano\second}$ and $\omega_c^\text{op}=\SI{4.5}{\giga\hertz}$, unless otherwise specified.
                Model parameters are identical to those used in the main text.
            }
        \end{figure}
        
        \begin{figure}[t]
            \centering
            \includegraphics[scale=1]{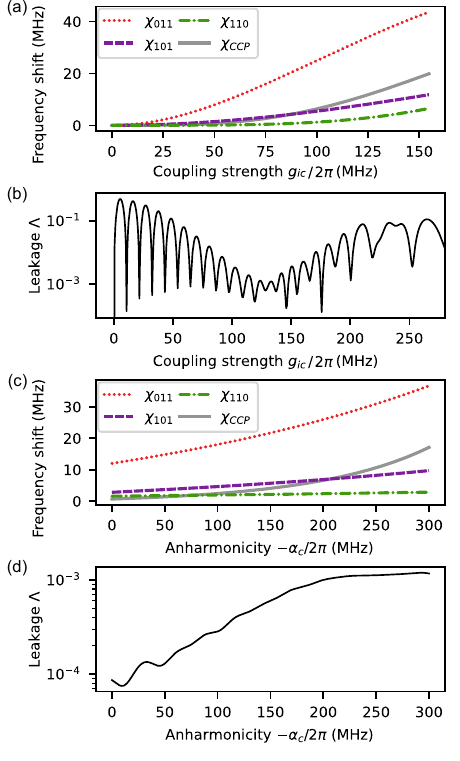}
            \caption{
                Frequency shifts and leakage as a function of coupling strength and anharmonicity $\alpha_c$. (a) Two-qubit energy shifts $\chi_{n_1 n_2 n_3}$ and the three-qubit energy shift $\chi_{CCP}$ as well as (b) leakage $\Lambda$ after a flux pulse as a function of the coupling strength $g_{ic}$. (c) Energy shifts and  (d) leakage as a function of the coupler anharmonicity $\alpha_c$.
                The energy shifts are calculated at $\omega_c/2\pi = \SI{4.5}{\giga\hertz}$.
                For the leakage simulation, a half-Gaussian-square pulse is used, with total duration $\tau = \SI{60}{\nano\second}$, rise time $\tau_R = \SI{5}{\nano\second}$ and $\omega_c^\text{op}=\SI{4.5}{\giga\hertz}$.
                Other model parameters are identical to those used in the main text.
            } \label{fig:simple_param_dep_Toff}
        \end{figure}

        \section{\texorpdfstring{\\* \vspace{2mm}}~Optimization of Pulse Parameters}
        \label{app:pulse_par_opt}
        To maximize the gate fidelity we investigate the effect of control parameters on the system dynamics, specifically, the accumulated phase during a flux pulse and the associated leakage.
        The flat-top pulse tuning the coupler from the idling to the interaction region is described by Eq.~\eqref{eq:half_Gaussian_square_pulse} with a rise-fall parameter $\tau_R=\SI{5}{\nano\second}$, a pulse length $\tau$ and the operation point $\omega_c^\text{op}$.
        Varying the pulse length $\tau$ linearly increases the accumulated phases $\phi_j$ [Eq.~\eqref{eq:cond_phases}], as shown in Fig.~\ref{fig:phase_leak_amp_t_plot}(a).
        Fitting this phase with a linear function provides the effective energy shifts $\chi_j$ used in Eq.~\eqref{eq:full_ccphase_lin_equation}, plus an offset given by the leading and trailing ramps.
        The population losses $\Lambda_{n_{1}n_{2}n_{3}}$ from the adiabatic states $\ket{n_{1}n_{2}n_{3}}$ remain approximately constant as a function of the pulse duration, with oscillations due to Landau-Zener-St\"{u}ckelberg interference~\cite{Shevchenko2010}, suggesting that leakage is introduced only during the ramps of the pulse and not while in the interaction region [see Fig.~\ref{fig:phase_leak_amp_t_plot}(b)].
        With this pulse the slope of the ramp and thus its adiabaticity can be tuned by the rise-fall parameter $\tau_R$. Individual pulses could be optimized to significantly suppress leakage by enforcing local adiabaticity, as well as utilizing destructive Landau-Zener-St\"{u}ckelberg interference~\cite{Rol2019, Negirneac2021}.
        Varying, instead, the operation point $\omega_c^\text{op}$ for a pulse with fixed duration $\tau = \SI{55}{\nano\second}$ will affect the magnitudes of the energy shifts so that lower $\omega_c^\text{op}$ leads to larger accumulated phases, as shown in Fig.~\ref{fig:phase_leak_amp_t_plot}(c).
        When choosing $\omega_c^\text{op}$ close to the four-photon avoided crossings $\ket{1,200}^0\leftrightarrow\ket{0,111}^0$ and $\ket{1,100}^0\leftrightarrow\ket{0,011}^0$ the involved states experience strong population losses [see Fig.~\ref{fig:phase_leak_amp_t_plot}(d)].
        Instead, an operation point detuned from these crossings results in an improved diabatic passage and thus reduced leakage.
        Indeed, the results presented in the main text use an operation point near Q$_3$ ($\omega_c^\text{op} \approx \omega_3 = \SI{4.5}{\giga\hertz}$) where we find a trade-off between low leakage and significant amounts of accumulated phase.
        Lowering the coupler frequency below this point leads to very strong hybridizations between states with excitations in the coupler and states with excitations in Q$_3$, which would experience more leakage.

        \section{\texorpdfstring{\\* \vspace{2mm}}~Optimization of Design Parameters}
        \label{app:des_par_opt}
        In optimizing design parameters we aim to maximize the strength of the energy shifts, which determines the gate speed, and minimize the amount of flux pulse leakage, which represents the dominant contribution of coherent errors.
        We limit our investigation to the coupling strengths $g_{ic}$ between qubits and coupler and the anharmonicity $\alpha_c$ of the tunable coupler, as these will determine the width and position of the avoided crossings relevant for the gate [see Appendix~\ref{app:full_energies}].

        We find that the conditional energy shifts $\chi_j$, as determined by Hamiltonian diagonalization, increase with increasing coupling strengths $g_{ic}$ due to a greater energy gap of the avoided crossing [see Fig.~\ref{fig:simple_param_dep_Toff}(a)].
        However, numerical simulation of the total final leakage $\Lambda=\Sigma_{n_1,n_2,n_3}\Lambda_{n_{1}n_{2}n_{3}}$ as a function of the coupling strengths $g_{ic}$ exhibits nonmonotonic behaviour, with oscillations due to Landau-Zener-St\"{u}ckelberg interference~\cite{Shevchenko2010}, and a stable minimum around $g_{ic} \sim \SI{120}{\mega\hertz}$ [see Fig.~\ref{fig:simple_param_dep_Toff}(b)].
        For smaller coupling strengths $g_{ic}$ the avoided crossings with excited coupler states become too narrow to be passed adiabatically and for larger coupling strengths $g_{ic}$ the avoided crossings that should be passed diabatically become too wide to do so.

        Increasing the magnitude of the anharmonicity of the coupler $\alpha_c$
        brings the three main avoided crossings introducing energy shifts closer together.
        Therefore, the energy shifts at an operation point, here $\omega_c^\text{op} \approx \SI{4.5}{\giga\hertz}$ increase as a function of $-\alpha_c$, as shown in Fig.~\ref{fig:simple_param_dep_Toff}(c).
        The stronger energy shifts come at the cost of more simultaneous hybridizations and frequency crowding, which increase the amount of leakage [Fig.~\ref{fig:simple_param_dep_Toff}(d)].
        Nonetheless, the introduced leakage is limited to approximately $10^{-3}$ and the choice of anharmonicity will be mostly guided by charge noise considerations for transmon qubits.

    \end{appendices}
    \bibliography{bibliography}

\end{document}